\documentclass{article}
\usepackage[square,numbers]{natbib}

\usepackage[final]{style}

\usepackage{style,times}
\usepackage{hyperref}
\usepackage{url}
\usepackage{graphicx}
\usepackage{booktabs} 
\usepackage{adjustbox}
\usepackage{colortbl} 
\usepackage{xcolor} 
\usepackage[skins]{tcolorbox}

\title{SpeechPy - A Library for Speech Processing and Recognition}

\author{Amirsina Torfi \\
  Department of Computer Science\\
  Virginia Tech\\
  \texttt{amirsina.torfi@gmail.com} \\
}

\begin{document}

\maketitle

\begin{abstract}
  SpeechPy is an open source Python package that contains speech preprocessing techniques, speech features, and important post-processing operations. It provides most frequent used speech features including MFCCs and filterbank energies alongside with the log-energy of filter-banks. The aim of the package is to provide researchers with a simple tool for speech feature extraction and processing purposes in applications such as Automatic Speech Recognition and Speaker Verification.
\end{abstract}

\section{Overview}

\begin{figure}[h]
\begin{center}
\includegraphics[scale=0.5]{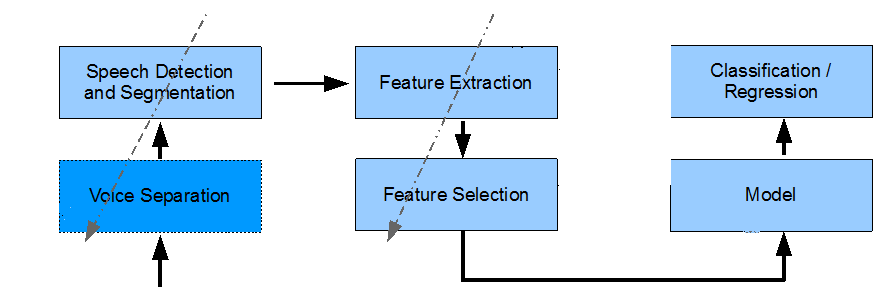}
\end{center}
\caption{Scheme of speech recognition system.}
\label{fig:speechscheme}
\end{figure}

Automatic Speech Recognition~(ASR) requires three main components for further analysis: Preprocessing, feature extraction, and post-processing. Feature extraction, in an abstract meaning, is extracting descriptive features from raw signal for speech classification purposes (Fig.~\ref{fig:speechscheme}). Due to the high dimensionality, the raw signal can be less informative compared to extracted higher level features. Feature extraction comes to our rescue for turning the high dimensional signal to a lower dimensional and yet more informative version of that for sound recognition and classification~\cite{furui1986speaker,guyon2008feature,hirsch2000aurora}.

Feature extraction, in essence, should be done considering the specific application at hand. For example, in ASR applications, the linguistic characteristics of the raw signal are of great importance and the other characteristics must be ignored~\cite{yu2016automatic,rabiner1993fundamentals}. On the other hand, in Speaker Recognition~(SR) task, solely voice-associated information must be contained in extracted feature~\cite{campbell1997speaker}. So the feature extraction goal is to extract the relevant feature from the raw signal and map it to a lower dimensional feature space. The problem of feature extraction has been investigated in pattern classification aimed at preventing the curse of dimensionality. There are some feature extraction approaches based on information theory \cite{torfi2017construction,shannon2001mathematical} applied to multimodal signals and demonstrated promising results~\cite{gurban2009information}.

The speech features can be categorized into two general types of acoustic and linguistic features. The former one is mainly related to non-verbal sounds and the later one is associated with ASR and SR systems for which verbal part has the major role. Perhaps one the most famous linguistic feature which is hard to beat is the Mel-Frequency Cepstral Coefficients~(MFCC). It uses speech raw frames in the range from 20ms to 40ms for having stationary characteristics~\cite{rabiner1993fundamentals}. MFCC is widely used for both ASR and SR tasks and more recently in the associated deep learning applications as the input to the network rather than directly feeding the signal~\cite{deng2013recent,lee2009unsupervised,yu2011improved}. With the advent of deep learning~\cite{lecun2015deep,torfi2018attention}, major improvements have been achieved by using deep neural networks rather than traditional methods for speech recognition applications~\cite{variani2014deep,hinton2012deep,liu2015deep}.

With the availability of free software for speech recognition such as VOICEBOX\footnote{\url{http://www.ee.ic.ac.uk/hp/staff/dmb/voicebox/voicebox.html}}, most of these softwares are Matlab-based which limits their reproducibility due to commercial issues. Another great package is PyAudioAnalysis~\cite{giannakopoulos2015pyaudioanalysis}, which is a comprehensive package developed in Python. However, the issue with PyAudioAnalysis is that its complexity and being too verbose for extracting simple features and it also lacks some important preprocessing and post-processing operations for its current version.

Considering the recent advent of deep learning in ASR and SR and the importance of the accurate speech feature extraction, here are the motivations behind SpeechPy package:

\begin{itemize}
\item Developing a free open source package which covers important preprocessing techniques, speech features, and post-processing operations required for ASR and SR applications.
\item A simple package with a minimum degree of complexity should be available for beginners.
\item A well-tested and continuously integrated package for future developments should be developed.

\end{itemize}

SpeechPy has been developed to satisfy the aforementioned needs. It contains the most important preprocessing and post-processing operations and a selection of frequently used speech features. The package is free and released as an open source software\footnote{\url{https://github.com/astorfi/speechpy}}. Continuous integration using for instant error check and validity of changes has been deployed for SpeechPy. Moreover, prior to the latest official release of SpeechPy, the package has successfully been utilized for research purposes~\cite{torfi20173d,torfi2017text}.

\section{Package Eco-system}

SpeechPy has been developed using Python language for its interface and backed as well. An empirical study demonstrated that Python as a scripting language, is more effective and productive than conventional languages\footnote{such as C and Java} for some programming problems and memory consumption is often "better than Java and not much worse than C or C++"~\cite{prechelt2000empirical}. We chose Python due to its simplicity and popularity. Third-party libraries are avoided except \textit{Numpy} and \textit{Scipy} for handling data and numeric computations.

\subsection{Complexity}
\begin{figure}[h]
\begin{center}
\includegraphics[scale=0.5]{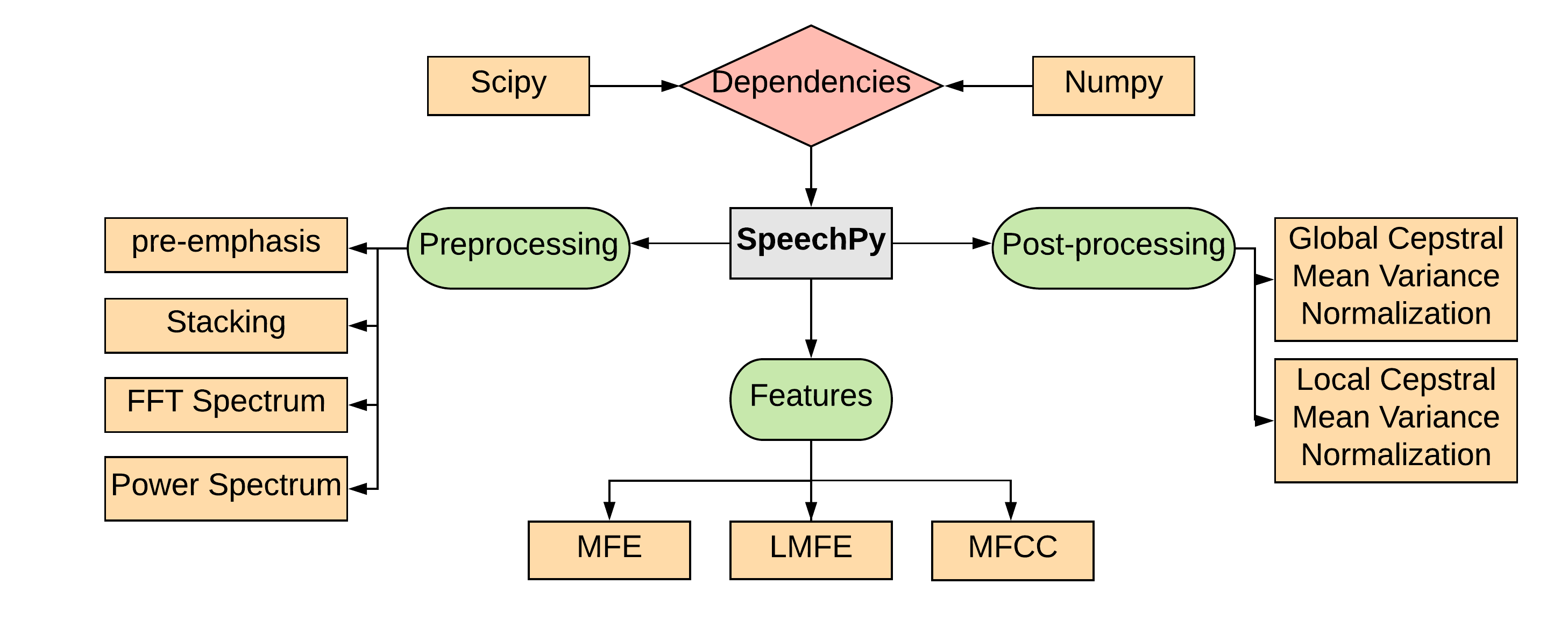}
\end{center}
\caption{A general view of the package.}
\label{fig:packageview}
\end{figure}

As the user should not and does not even need to manipulate the internal package structure, object-oriented programming is mostly used for package development which provides easier interface for the user with a sacrifice to the simplicity of the code. However, the internal code complexity of the package does not affect the user experience since the modules can easily be called with the associated arguments. SpeechPy is a library with a collection of sub-modules. The general scheme of the package is provided in Fig.~\ref{fig:packageview}.

\subsection{Code Style and Documentation}
SpeechPy is constructed based on PEP 8 style guide for Python codes. Moreover, it is extensively documented using the formatted docstrings and Sphinx\footnote{\url{http://www.sphinx-doc.org}} for further automatic modifications to the document in case of changing internal modules. The full documentation of the project will be generated in HTML and PDF format using Sphinx and is hosted online. The official releases of the project are hosted on the Zenodo as well\footnote{\url{https://zenodo.org/record/810391}} \cite{torfispeechpy}.

\subsection{Continuous Testing and Extensibility}

\begin{figure}[h]
\begin{center}
\includegraphics[scale=0.4]{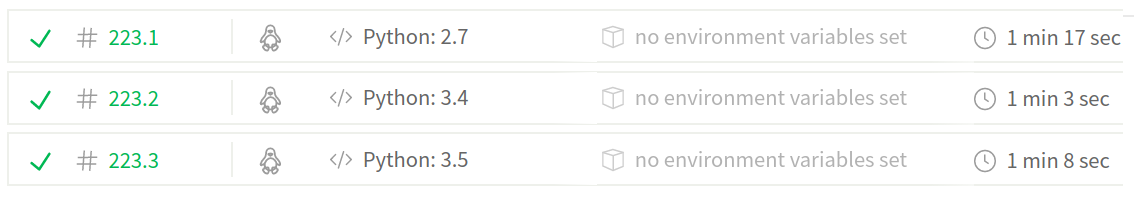}
\end{center}
\caption{Travic CI web interface after testing SpeechPy against a new change.}
\label{fig:travisci}
\end{figure}

The output of each function has been evaluated as well using different tests as opposed to the other existing standard packages. For continuous testing, the code is hosted on GitHub and integrated with Travis CI. Each modification to the code must pass the unit tests defined for the continuous integration. This will ensure the package does not break with unadapted code scripts. However, the validity of the modifications should always be investigated with the owner or authorized collaborators of the project. The code will be tested at each time of modification for Python versions \textit{"2.7"}, \textit{"3.4"} and \textit{"3.5"}. In the future, these versions are subject to change. The Travis CI interface is depicted in Fig.~\ref{fig:travisci}.

\section{Availability}

\subsection*{Operating system}

Tested on Ubuntu 14.04 and 16.04 LTS Linux, Apple Mac OS X 10.9.5 , and Microsoft Windows 7 \& 10. We expect that SpeechPy works on any distribution as long as Python and the package dependencies are installed.

\subsection*{Programming language}

The package has been tested Python 2.7, 3.4 and 3.5. However, using Python 3.5 is suggested.

\subsection*{Additional system requirements \& dependencies}

SpeechPy is a light package and small computational power would be enough for running it. Although the speed of the execution is totally dependent to the system architecture. The dependencies are as follows:
\begin{itemize}
\item Numpy
\item SciPy
\end{itemize}

\section{Aknowledgement}

This work has been completed in part with computational resources provided by the West Virginia
University and is based upon a work supported by the Center for Identification Technology Research~(CITeR) and the National Science Foundation~(NSF) under Grant \#1650474.

%
%

\bibliographystyle{unsrtnat}
\bibliography{ref}
\end{document}